\documentclass[sigconf]{acmart}
\AtBeginDocument{%
  }

\setcopyright{none}
\copyrightyear{}
\acmYear{}
\acmDOI{}

\acmConference[]{}{}{}
\acmISBN{}

\renewcommand\footnotetextcopyrightpermission[1]{}

\usepackage{color}
\usepackage{amsmath}
\usepackage{ascmac}

\usepackage{todonotes}
\usepackage{booktabs} % For formal tables
\usepackage{subfig}
\usepackage{colortbl}
\usepackage{xcolor}
\definecolor{myb}{rgb}{0.875,0.95,1}
\definecolor{myp}{rgb}{1.0, 0.89,0.879}
\usepackage{amsthm}
\begin{document}
\pagestyle{empty}

%%
%% The "title" command has an optional parameter,
%% allowing the author to define a "short title" to be used in page headers.
\title{Impact of Tone-Aware Explanations in Recommender Systems}

%%
%% The "author" command and its associated commands are used to define
%% the authors and their affiliations.
%% Of note is the shared affiliation of the first two authors, and the
%% "authornote" and "authornotemark" commands
%% used to denote shared contribution to the research.
\author{Ayano Okoso}
\affiliation{%
  \institution{Toyota Central R\&D Labs., Inc. }
  \streetaddress{41-1 Yokomichi Nagakute}
  \city{Aichi}
  \country{Japan}}
\email{okoso@mosk.tytlabs.co.jp}

\author{Keisuke Otaki}
\affiliation{%
  \institution{Toyota Central R\&D Labs., Inc. }
  \streetaddress{41-1 Yokomichi Nagakute}
  \city{Aichi}
  \country{Japan}}
\email{otaki@mosk.tytlabs.co.jp}

\author{Satoshi Koide}
\affiliation{%
  \institution{Toyota Central R\&D Labs., Inc. }
  \streetaddress{41-1 Yokomichi Nagakute}
  \city{Aichi}
  \country{Japan}}
\email{koide@mosk.tytlabs.co.jp}

\author{Yukino Baba}
\affiliation{%
  \institution{The University of Tokyo}
  \streetaddress{3-8-1 Komaba Meguro}
  \city{Tokyo}
  \country{Japan}}
\email{yukino-baba@g.ecc.u-tokyo.ac.jp}

%%
%% By default, the full list of authors will be used in the page
%% headers. Often, this list is too long, and will overlap
%% other information printed in the page headers. This command allows
%% the author to define a more concise list
%% of authors' names for this purpose.
\renewcommand{\shortauthors}{Okoso et al.}

%%
%% The abstract is a short summary of the work to be presented in the
%% article.
\begin{abstract}
In recommender systems, the presentation of explanations plays a crucial role in supporting users' decision-making processes. 
Although numerous existing studies have focused on the effects (transparency or persuasiveness) of explanation content, explanation expression is largely overlooked.
Tone, such as formal and humorous, is directly linked to expressiveness and is an important element in human communication.
However, studies on the impact of tone on explanations within the context of recommender systems are insufficient.
Therefore, this study investigates the effect of explanation tones through an online user study from three aspects: perceived effects, domain differences, and user attributes.
We create a dataset using a large language model to generate fictional items and explanations with various tones in the domain of movies, hotels, and home products.
Collected data analysis reveals different perceived effects of tones depending on the domains.
Moreover, user attributes such as age and personality traits are found to influence the impact of tone.
This research underscores the critical role of tones in explanations within recommender systems, suggesting that attention to tone can enhance user experience.
\end{abstract}

%%
%% The code below is generated by the tool at http://dl.acm.org/ccs.cfm.
%% Please copy and paste the code instead of the example below.
%%


%%
%% Keywords. The author(s) should pick words that accurately describe
%% the work being presented. Separate the keywords with commas.
%% A "teaser" image appears between the author and affiliation
%% information and the body of the document, and typically spans the
%% page.
% \begin{teaserfigure}
%   \includegraphics[width=\textwidth]{sampleteaser}
%   \caption{Seattle Mariners at Spring Training, 2010.}
%   \Description{Enjoying the baseball game from the third-base
%   seats. Ichiro Suzuki preparing to bat.}
%   \label{fig:teaser}
% \end{teaserfigure}

% \received{20 February 2007}
% \received[revised]{12 March 2009}
% \received[accepted]{5 June 2009}

%%
%% This command processes the author and affiliation and title
%% information and builds the first part of the formatted document.
\maketitle

\section{Introduction}\label{sec:intro}

Although recommender systems~\cite{resnick1997recommender} are essential tools for supporting users’ decision-making in several daily activities, they are often considered \textit{black boxes}.
This perception arises because it is usually unclear what information is used for making recommendations, what the underlying mechanism is, and how the recommendations align with users' preferences or needs.
Without understanding such backgrounds, users face difficulty in accepting suggestions even when systems accurately predict users' preferences and recommend items~\cite{sinha2002role}.

To mitigate the black box problem for users, providing explanations in recommender systems has attracted attention in recent years~\cite{knijnenburg2012explaining,zhang2020explainable}.
Explanations primarily target two aspects: model and result~\cite{miller2019explanation,zhang2020explainable}.
Model explanations inform users about the logic to predict recommended items and improve transparency ~\cite{zhang2020explainable}.
Conversely, result explanations, also known as \textit{justification explanation}~\cite{miller2019explanation,musto2019justifying}, do not accurately explain recommendation model mechanisms.
Instead, they justify the recommendation results by providing relevance to the users' past purchase history or indicating aspects the users would likely prefer in the recommended items. 
Further, justification explanation improves users’ perceived effectiveness and satisfaction~\cite{vig2009tagsplanations,musto2019justifying}.
Thus, providing explanations enhances user trust in recommender systems and contributes to users’ decision-making. 
Herein, we focus on the textual justification explanation and refer to it as \textit{explanation} for simplicity.

Although several existing studies have focused on explanation content, the \textit{expression} of explanations has been largely overlooked.
Conversely, such explanations have been typically conveyed using uniform or fixed phrases.
For example, a collaborative-based explanation~\cite{herlocker2000explaining,cosley2003seeing}, which recommends based on similarities between users' preferences or behaviors, states,  ``Users with preferences similar to yours also purchased this item.''
A feature-based explanation~\cite{vig2009tagsplanations,hou2019explainable}, focusing on specific product attributes or features that align with users' interests or requirements, describes, ``You might be interested in a battery on which this product performs well.'' 
Thus, several studies have employed standardized explanations to investigate the effects of explanation content.
However, to the best of our knowledge, the effect of expressiveness is yet to be explored.

\textit{Tone}~\cite{jin2022deep}, including formal and humorous, is directly linked to expressiveness and is a central element in human communication.
While recommending something to others, we usually adjust the content along with our tone to suit the recipient~\cite{giles2007communication,niederhoffer2002linguistic}.
For example, when we recommend movies to others, we often vary our wordings and expressions depending on the recipient’s personality or relationship.
Some people prefer an emotional tone, while others prefer a restrained tone even with the same content.
Additionally, a light tone is more appropriate when the explanation subject involves entertainment, while a formal tone is more suitable for practical or highly specialized topics.
Although tone can enrich the expressiveness of explanations in recommender systems, its impact has not been explored.

Therefore, this study investigates the impact of tone on item explanations within the context of recommender systems, guided by the following three research topics.\\
\textbf{Perceived effects of tones.}
In context of recommender systems, seven explanation goals include: transparency, scrutability, trust, effectiveness, persuasiveness, efficiency, and satisfaction~\cite{tintarev2012evaluating}.
Previous studies have investigated the goals affected by explanation content and presentation format (i.e., textual or visual)~\cite{zhang2014explicit,wang2018explainable,kouki2019personalized}.
However, which of the seven goals is influenced by tone is yet to be clarified.
Thus, we investigate the impact of tone on the seven goals in this study.
Additionally, the goals are based on the information contained in the explanations. 
Even with the same information, different ways of expression can change users' impressions of recommended items. 
Therefore, we introduce a new metric defined as the ability to arouse interest in recommended items, referred to as \textit{appeal}, thereby forming a set of eight effects.
While appeal may have indirect relations with effectiveness, satisfaction, and persuasiveness, we consider it an independent effect due to its significance as an element of expressiveness. 
Among the eight effects, we aim to investigate those influenced by tone, seeking a deeper understanding of expressiveness in explanations.
Consequently, we pose the following research question: \textit{How does the tone of explanations influence perceived effects?}\\
\textbf{Relation between the effects of tones and domains.} 
The impact of tones can vary based on domains.
For example, a tone can strongly influence perceived effects in short-term entertainment domains such as movies and music.
This might be because users perceive low risks associated with consuming items in these domains, leading users to value impressions and empathy evoked by explanations more than the content.
Contrarily, in practical domains where items are used over a longer period, including home products, tonal influence can be less significant.
Since such items are intended for long-term use and often involve higher costs, they represent higher consumption risks.
Consequently, users may prioritize specific and detailed information in explanations when deciding to accept recommended items.
To address the hypothesis, we investigate tonal effects across three different domains, considering entertainment and consumption duration as criteria: movies (short-term, entertainment), hotels (mid-term, entertainment/practical), and home products (long-term, practical).
Therefore, we pose the following research question: \textit{How do the perceived effects of tones vary across domains?}\\

\textbf{Relation between the effects of tones and user attributes.}
User attributes, including age, gender, personality, and expertise, influence interaction between recommender systems and users~\cite{dhelim2022survey}. 
Thus, we hypothesize that tonal effects depend on user attributes.
For example, users with certain personality traits can be less susceptible to the influence of tone.
To test the hypothesis, we investigate correlation between user attributes and tonal effects in explanations.
Our research question is: \textit{How do the perceived effects of tones depend on user attributes?}

In this study, we conduct an online user survey to investigate tonal effects considering abovementioned perspectives.
Specifically, we develop a dataset for our survey by generating fictional items and explanations in various tones using a large language model (LLM) for \textit{movie}, \textit{hotel}, and home product (\textit{product} for simplicity) domains (Sec.~\ref{subsec:data}).
We show participants the items with pairs of explanations in different tones and ask them to perform a pairwise comparative evaluation against the eight explanation effects (Sec.~\ref{sec:user_study}).
Using the collected data, we analyze tonal influence on each goal and the relationship between tonal effects, domains, and user attributes (Sec.~\ref{sec:result}).
As a result, the contributions of this study include the following findings.
\begin{description}
    \item [Perceived effects of tones:] Although differences exist depending on domains and user attributes, significant tonal influences are observed across 10 metrics based on the eight effects: transparency, scrutability, trust (competence, benevolence, integrity), effectiveness, persuasiveness, efficiency, satisfaction, and appeal. Explanations with rich expressions positively affect users in various metrics more than plain explanations.  
    \item [Domain-specific trends:] We observe variations in the impact of tone across domains. Almost all metrics are significantly influenced by tones in the hotel domain, while the influences in movie and product domains remain limited.
    \item [Influence of user attributes:] We observe non-uniformity in the impact of tones across all users; it varies with user attributes including age and Big Five personality traits (e.g., extroversion). Thus, personalizing tones based on user profiles can change user perception of recommender systems.
\end{description}

\section{Related Work}\label{sec:related}

\subsection{Explanations in Recommender Systems}
The need for presenting explanations in recommender systems is increasing in both research and industrial fields to support user decision-making~\cite{zhang2020explainable,cramer2008effects}. Explanations can be delivered in various formats, including natural language~\cite{chang2016crowd,kouki2019personalized}, charts~\cite{hou2019explainableRadar,herlocker2000explaining}, or visual annotations on images~\cite{lin2019explainable,hou2019explainable}.
Zhang et al.~\cite{zhang2020explainable} categorized explanation contents into feature-based explanations~\cite{vig2009tagsplanations,hou2019explainable}, emphasizing item features or aspects that users prefer; collaborative filtering~\cite{herlocker2000explaining,cosley2003seeing}, focusing on similarities between users' preferences or items; and social relation-based explanations~\cite{park2017uniwalk,QUIJANOSANCHEZ201736}, highlighting connections in social networks.

Previous studies have mainly focused on how the content and format of explanations impact the seven goals of explanations, such as transparency and persuasiveness~\cite{tintarev2012evaluating}.
Herlocker et al.~\cite{herlocker2000explaining} evaluated 21 types of explanations and demonstrated the enhancement of persuasiveness of recommender systems by specific explanation contents and formats.
Kouki et al.~\cite{kouki2019personalized} evaluated different explanation contents and formats for a music recommender system.
They found textual feature-based explanations to be more persuasive than collaboration- or social relation-based explanations.
Furthermore, they revealed that the preferred explanation content varied with personality traits.
Hirschmeier et al.~\cite{Hirschmeier2020} clarified that users feel the transparency and trust in a radio domain even without referring to explanations as long as they know the availability at any time.
Additionally, Tran et al.~\cite{tran2023user} investigated the need for explanations across broader domains.
They found that users tended to read explanations when dissatisfied with recommended items, in domains requiring significant decisions, and when satisfied with items in more casual domains.
Further, they revealed that preferred explanation content was depended on domains, and the needs do not vary based on user attributes.

While previous studies on explanations have comprehensively covered explanation content, format, and requirements, the explanations used have been uniform and the effects of expressiveness (i.e., tone) have not been explored.
Therefore, our study investigates tonal effects in explanations and their correlations with user attributes across multiple domains. 
We aim to provide insights into the role of tones as a new element in explanations for recommender systems.

\begin{table*}[t]
\centering
\small
\caption{Input and output for each domain.}
\label{tb:item_inout}
\begin{tabular}{p{1.5cm}p{7cm}p{7cm}}
\toprule
Domain & Input & Output \\
\midrule
Movies & Genre (Drama, Comedy, Action, Sci-Fi, Thriller) & 
  Title, Synopsis (From 50 to 100 words), Poster image,\\ 
  \midrule
Hotels& Hotel class (5-class), & 
  Hotel name, Rooms, Services and Facilities, \\
 & Location (city, forest, sea) & Dining, Atmosphere and Hotel Style, Hotel image \\
  \midrule
Products & Item (Smartphone, Digital Camera, Speaker, & Item Name, Specifications, Features,\\
 & Coffee Maker, Toaster), Price Range (3 classes) & Functions, Design, Product image \\
 \bottomrule
\end{tabular}
\end{table*}

\subsection{Effects of Tones}

Tone is a subject of research in a wide range of communities, including linguistics~\cite{biber2019register,yule2022study}, psychology~\cite{niederhoffer2002linguistic,bonvillain2019language,packard2021concrete}, and computer science~\cite{shen2017style,jin2022deep}.
For this study, we define \textit{tone} as the expressiveness or style of written or spoken language, and categorize it into styles including formal or humorous.
The definition is rooted as \textit{text style} in the field of natural language processing~\cite{jin2022deep}, and encompasses aspects such as lexical choice and grammatical structure of texts.

Human communications involve content along with verbal elements, including word choices, tones, and inflections, and non-verbal elements, including facial expressions and body language~\cite{bonvillain2019language,scherer2003vocal}. 
Particularly, tone contributes to facilitate communication and we adjust tonal suitably in our daily lives depending on situations and our relationship with others~\cite{niederhoffer2002linguistic}. 
Furthermore, tone is effective when engaging or persuading other people~\cite{packard2021concrete, ludwig2013more}.

For human and artificial intelligence (AI) interactions, research on AI interventions and advices to support users' decision-making have gained attention~\cite{tolmeijer2022capable,hou2023should}.
Regarding the delivery method of interventions, existing literature~\cite{gilad2021effects} shows that users tend to prioritize warmth over competence level in AI advice.
In contrast, the literature~\cite{kahr2023seems} suggests that the human-like quality of explanations does not significantly influence the perceived reliability of the AI models.

Since such studies are for AI assistants that involve explicit intervention in the user's decision-making, they may not necessarily apply to recommender systems where interventions are less direct or explicit.
Previous studies on recommender systems have mainly concentrated on presenting the type and format of information~\cite{zhang2014explicit,wang2018explainable,kouki2019personalized}, without fully exploring the impacts of expressiveness. 
The study~\cite{Kunkel2019trust} has reported an increase in user satisfaction and trust when more expressive explanations are provided.
Thus, we focus on the expressiveness of explanations (i.e., tone) in recommender systems and explore their perceived impacts.

\section{User Study}\label{sec:user_study}

\subsection{Research Questions}\label{subsec:RQ}
We expect the impacts of tones to depend on domains and user attributes, as mentioned in Sec.~\ref{sec:intro}.
To discuss tonal impacts from the perspective of domains and user attributes, we raised three research questions as follows.
\begin{description}
    \item[RQ1]: How does the tone of explanations influence perceived effects regardless of domains and user attributes?
    \item[RQ2]: How do the perceived effects of tones vary across domains regardless of user attributes?
    \item[RQ3]: How do the perceived effects of tones depend on user attributes?
\end{description}
By answering these research questions, we aim to determine how the tone of explanations operates across various domains and how the impact of tone is affected by certain user attributes. Our research is the first attempt to reveal the effect of expressiveness in explanations for recommender systems, highlighting the importance of carefully handling the new element in developing future explainable recommendations.

\subsection{Data Preparation}\label{subsec:data}
For user study, we need to prepare datasets of diverse explanations converted to various tones.
We employed item descriptions as explanations that conveyed item features and attractiveness because other types of explanation, including collaboration-based, tend to be short with little room for tone conversion.
Moreover, they depend on the accuracy of capturing users’ preferences and can influence their perceptions.

We prepared datasets based on fictional data sources for three domains (movies, hotels, and products) inspired by an existing literature~\cite{tran2023user}.
Although various datasets (e.g., IMDb~\cite{maas-EtAl:2011:ACL-HLT2011}, TripAdvisor~\cite{datasetTripAdvisor}, and Amazon review dataset~\cite{ni2019justifying}) are available, we prepared data sources due to copyright-related concerns and the lack of suitable descriptions.
Specifically, existing datasets have a variety of user reviews, but they do not include descriptions of item features and attractiveness by item suppliers, motivating us to use fictional datasets\footnote{the fictitious dataset can be shared upon request}.
Additionally, fictional data sources can exclude checks related to participants in a user study having known the item.
Thus, we can eliminate the possibility of preconceived notions and biases arising from knowing the item in advance.
In the following sections, we explain the steps to prepare fictional data sources for three domains and then describe the tones covered in this study.

\subsubsection{Input and Output for Creating Data Sources}\label{subsubsec:inout}

We created 15 fictional items for each domain using LLM.
We set inputs and requirements to generate corresponding features of fictional items.
Table~\ref{tb:item_inout} shows the inputs and outputs for each domain.
In the movie domain, we input each of five genres and created three movies per genre, generating 15 movies. The output included title, plot, and poster image.
In the hotel domain, we input each of five hotel classes and three locations.
We generated hotel information for each combination, resulting in a total of 15 profiles.
The output included name, rooms, service and facility, food, atmosphere and style, and hotel image.
In the product domain, we input each of five item types and three price ranges.
We generated fictitious products for each combination, resulting in a total of 15 products.
The output included name, specifications, features, functions, design, and product image.
We set the aforementioned types of inputs and outputs referencing real services (e.g., IMDb~\footnote{https://www.imdb.com/}, hotels.com~\footnote{https://www.hoteles.com/}, and Amazon~\footnote{https://www.amazon.com/}).

\begin{table*}[t]
\centering
\small
\caption{Examples of generated fictional items for each domain. Each input includes (Movies) Genre: comedy, (Hotels) Location: forest, Class: 5, and (Products) Item: toaster, Price range: 2.}
\label{tb:item_example}
\begin{tabular}{p{5.2cm}p{6.5cm}p{5.0cm}}
\toprule
Movies & Hotels & Products \\
\midrule
\includegraphics[width=0.15\textwidth, height=3.5cm]{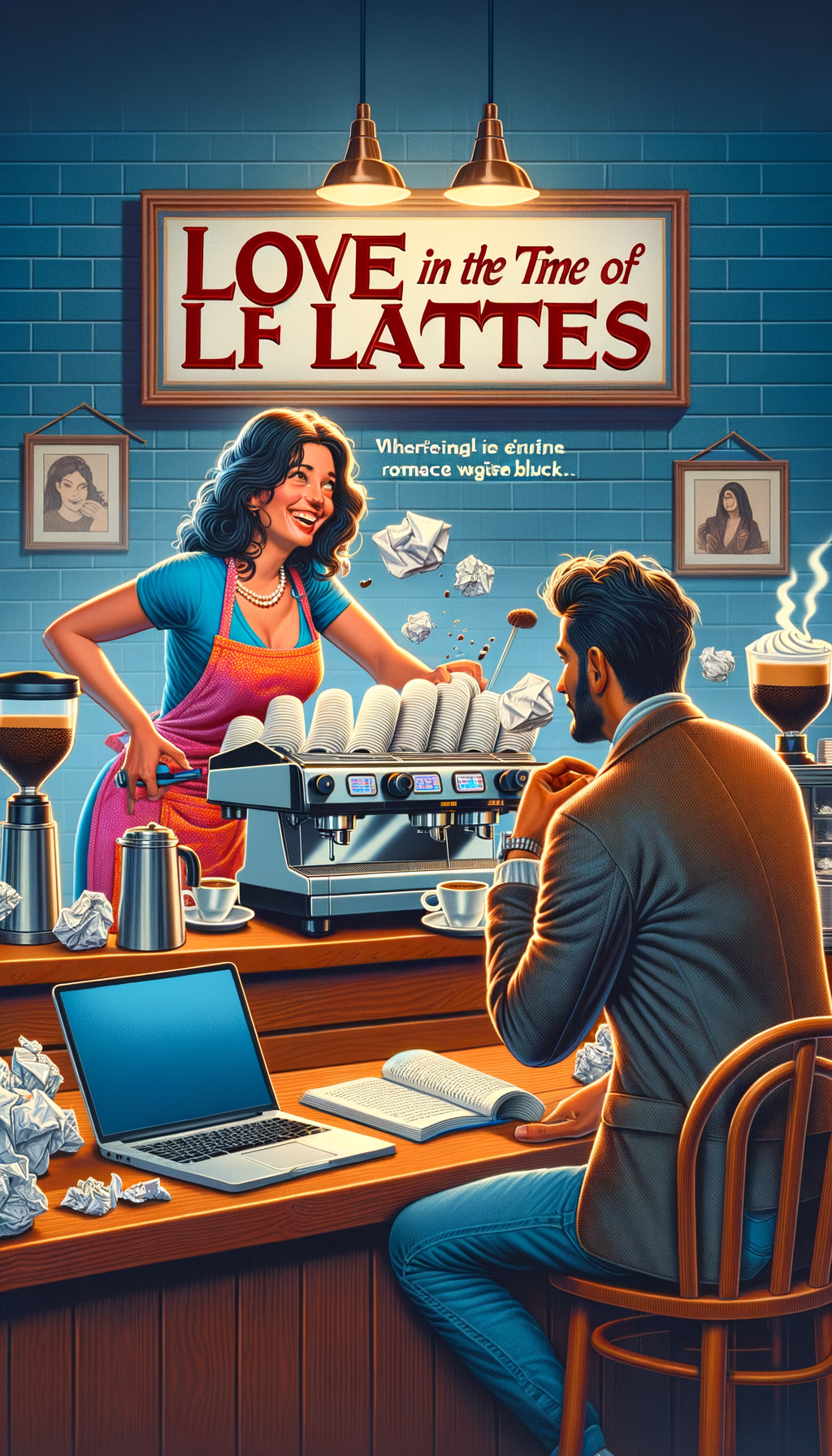} & \includegraphics[width=0.20\textwidth]{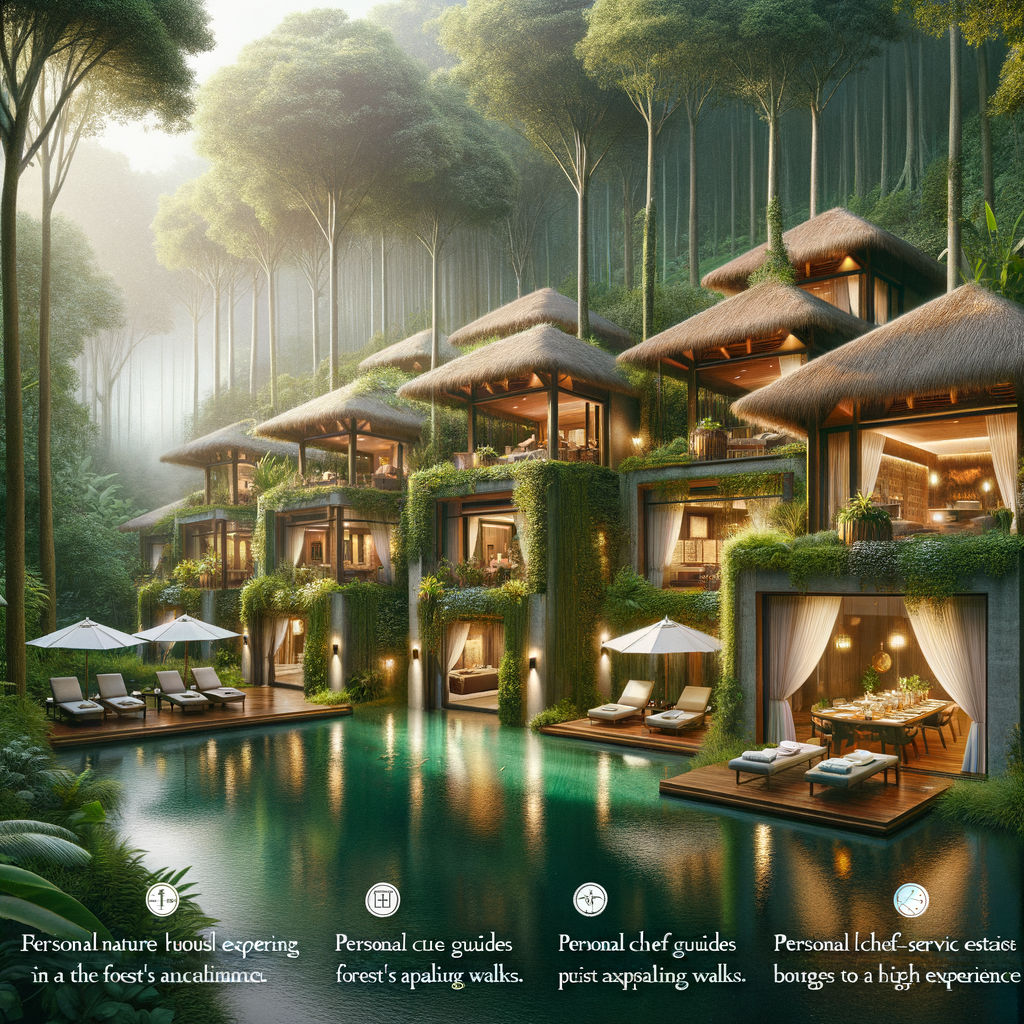} & \includegraphics[width=0.20\textwidth]{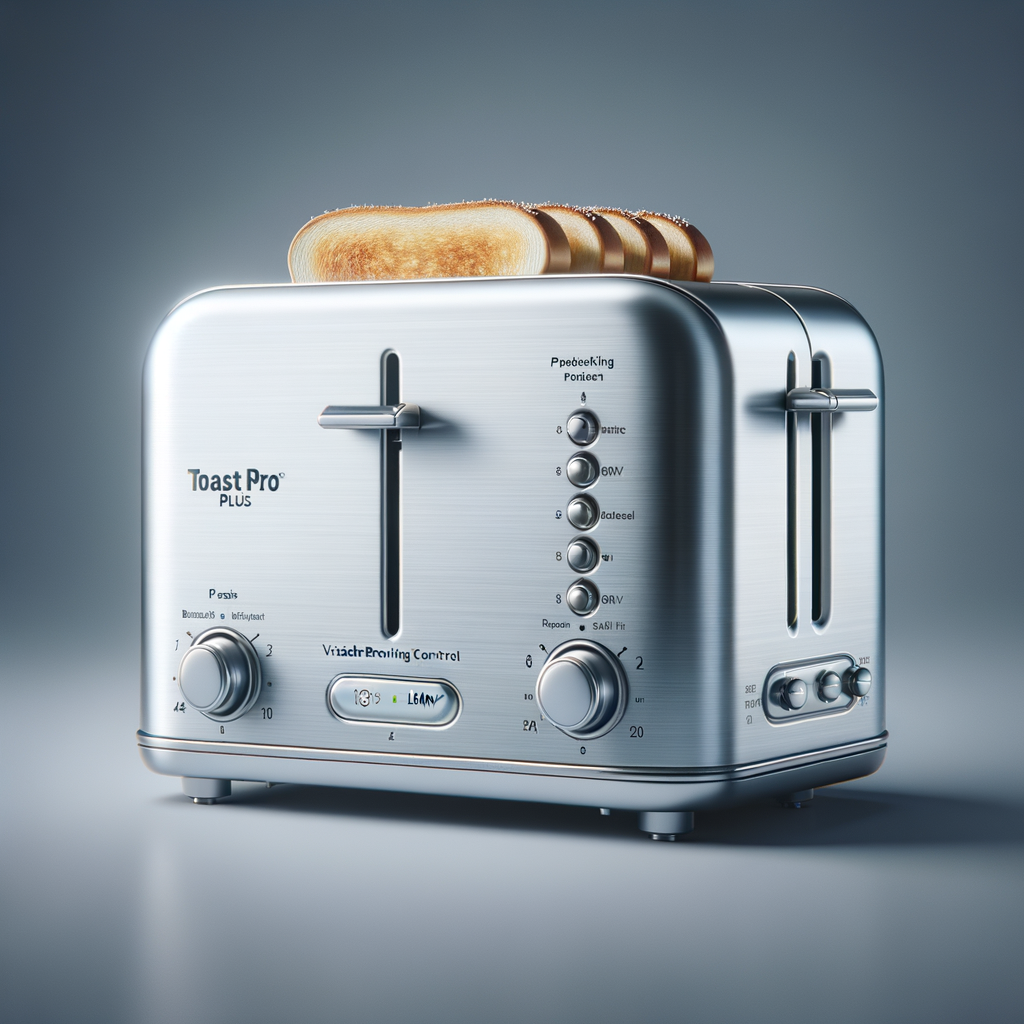} \\
\textbf{Title}: ``Love in the Time of Lattes''  & \textbf{Name}: Enchanted Forest Sanctuary & \textbf{Name}: ToastPro Plus \\
\textbf{Plot}: A tongue-in-cheek romantic comedy &  \textbf{Rooms}: 30 & \textbf{Spec.}: 4 slice capacity, 1500W power \\
where a barista falls for a regular customer & \textbf{Services and Facilities}: Private villas,& \textbf{Features}: Variable browning control, \\
who happens to be a romance novelist&  personal nature guides & Defrost function\\
suffering from writer's block.& \textbf{Dining}: Personal chef service &\textbf{Functions}: Toasting, Bagel toasting
 \\
& \textbf{Atmosphere and Hotel Style}: Ultra-luxurious, Private& \textbf{Design}: Sleek design with silver finish\\
\bottomrule
\end{tabular}
\end{table*}

\subsubsection{Creation of Fictional Items Using LLM}\label{subsubsec:item}

We used the application programming interface (API) of OpenAI~\footnote{https://openai.com/blog/openai-api} to generate fictional items: DALLE-2.0 for generating item images and GPT-4 for other item information.
We used the following common prompt to generate item information.
\begin{itembox}[l]{Prompt: Generating Item Information}
Generate fictional \texttt{[A]} based on the following input. The input and output are as follows: \\
Input: \texttt{[B]} \\
Output: \texttt{[C]} \\
Requirements: \texttt{[D]}
\end{itembox}
\texttt{[A]} contains \textit{movies}, \textit{hotels}, or \textit{products} corresponding to the domain.
\texttt{[B]} and \texttt{[C]} contain inputs and outputs excluding the image shown in Table~\ref{tb:item_inout}, respectively.
\texttt{[D]} depends on the domains as follows.\\
\textbf{Movies:} \textit{The input genres can be one of five options (Drama, Comedy, Action, Sci-Fi, Thriller). Generate 15 fictional movies, with three movies for each input genre.}\\
\textbf{Hotels:} \textit{The hotel class ranges from 1 to 5 (1 is low, 5 is high), and the location can be one of three options (city, forest, sea). There are 15 possible input combinations, so generate 15 diverse fictitious hotels. \textbackslash n
Set the hotel's price range based on the hotel class and location. Price range 1 represents budget-friendly hotels, and price range 5 represents high-end hotels.}\\
\textbf{Products:} \textit{Generate 15 fictional items, with three price ranges for each input item. \textbackslash n
Price range 1 represents budget-friendly items, and price range 3 represents high-end items.}

Similarly, we used the following common prompt to generate item images.
\begin{itembox}[l]{Prompt: Generating Item Image}
Generate a \texttt{[A]} image for the fictional \texttt{[B]} of the following input. \\ 
Input: \texttt{[C]}
\end{itembox}
\texttt{[A]} contains \textit{poster}, \textit{fictitious hotel}, and \textit{realistic photographic} corresponding to the movie, hotel, and product domains, respectively.
\texttt{[B]} includes movie, hotel, and product corresponding to the domains.
\texttt{[C]} contains fictional item information with inputs and outputs shown in Table~\ref{tb:item_inout}, excluding images.
We added the following requirements for hotel images as the focus of images varies in the hotel domain:
\textit{Choose randomly one of the following for the image subject: Exterior, Room, Interior, Facilities, or Dining.
The image should have a realistic photographic style.}

Table~\ref{tb:item_example} lists examples of item information and images generated following the aforementioned procedure. 
Note that we set the images in the movie domain as vertical, while those in the other domains are as square.

\begin{table*}[t]
\small
\centering
    \caption{Characteristics of each tone.}
    \label{tb:tone_chara}
    \begin{tabular}{ll}
    \toprule
    Tone & Characteristics\\
    \midrule
    Neutral (Neu.) & \small\begin{tabular}{@{}l@{}}
    Unbiased, objective, and devoid of emotional expression. \\
    Stick to factual information, avoid emotional language or personal opinions.\end{tabular}\\ \midrule
    Attractive (Att.) & \small\begin{tabular}{@{}l@{}}
    Captivating, engaging, and appealing to the senses or emotions. \\
    Use vivid adjectives, descriptive language, and storytelling techniques to make the sentence more interesting and engaging.\end{tabular}\\ \midrule
    Humorous (Hum.) & \small\begin{tabular}{@{}l@{}}
    Funny, witty, and intended to provoke laughter or amusement.\\
    Incorporate puns, wordplay, exaggerations, or unexpected twists to create humor.\end{tabular}\\ \midrule
    Romantic (Rom.) & \small\begin{tabular}{@{}l@{}}
    Expressive of love, passion, and deep affection.\\
    Use romantic imagery, metaphors, and poetic language to convey feelings of love and longing.\end{tabular}\\ \midrule
    Formal (For.) & \small\begin{tabular}{@{}l@{}}
    Polite, respectful, and adhering to established conventions of language.\\
    Avoid contractions, use proper titles, and maintain a professional and respectful tone.\end{tabular}\\ \midrule
    Simple (Sim.)  & \small\begin{tabular}{@{}l@{}}
    Simple, clear, and easily understandable by children.\\
    Use straightforward language, short sentences, and avoid complex vocabulary or concepts. Explain any unfamiliar terms.\end{tabular}\\
    \bottomrule
    \end{tabular}
\end{table*}

\subsubsection{Generation of Fictional Explanations by LLM}\label{subsubsec:description}

We generated 10 explanations per item.
We employed two types of descriptions that conveyed item features or attractiveness: descriptions written by item suppliers and user-review style descriptions.
The difference involved in their contents of users’ personal experiences.
The reviews could be regarded as explanations as they provided users with detailed information and suggestions about the item, potentially leading to their decision on item consumption~\cite{chen2018neural}.
We generated five fictional descriptions written by item suppliers and five fictional review-style descriptions using GPT-4. We used the following parametrized prompt.

\begin{itembox}[l]{Prompt: Generating Item Descriptions}
Generate five descriptions for the fictional \texttt{[A]} based on the following input. \\
Input: \texttt{[B]}\\
Requirements: \\
Each description should be between 50 to 150 words.\\
Each description should randomly include at least one of the following aspects: \texttt{[C]}.\\
\texttt{[D]}
\end{itembox}
\texttt{[A]} contains \textit{movies}, \textit{hotels}, or \textit{products} according to the target domain.
\texttt{[B]} contains fictional item information with inputs and outputs shown in Table~\ref{tb:item_inout}, excluding images.
\texttt{[C]} contains the candidates of aspects in descriptions for each domain: (movies) plot, theme, genre, characters, direction, cast, director, and awards; (hotel) locations, rooms, service and facility, dining, atmosphere and style; (product) features, functions, and designs.
\texttt{[D]} is defined depending on description types as follows.\\
\textbf{Descriptions by suppliers}: \textit{Each description assumes that the item supplier conveys the item's attractiveness to the guests.}\\
\textbf{Review-style descriptions}: \textit{Each description should include personal user experiences.}

\subsubsection{Tone Conversion by LLM}\label{subsubsec:tone}

We converted each description into six types of tones.
Table~\ref{tb:tone_chara} lists the tones considered for this study. 
We addressed major axes in the previous study~\cite{jin2022deep}: formality, humor and romance, simplicity, and engagingness.
As a result, we chose six tones: neutral, formal, humorous, romantic, attractive, and simple. We specified ``simple as if speaking to 10-year-old children'' as a guide for simplicity.
Note that these tones are not complementary.
We used GPT-4 to convert descriptions to each tone. The prompt is as follows.
\begin{itembox}[l]{Prompt: Converting Descriptions into Target Tone}
Convert the following input text to a/an \texttt{[A]} tone.
\texttt{[A]} tone is characterized as follows: \\ \texttt{[B]}\\
The converted text should be less than 150 words. \\
Input: \texttt{[C]}
\end{itembox}
\texttt{[A]} and \texttt{[B]} contain one of the six tones and the characteristics are listed in Table~\ref{tb:tone_chara}, respectively.
\texttt{[C]} is the generated descriptions in the previous subsection. Table~\ref{tb:tone_example} lists examples of tone converted descriptions.

\begin{table*}[t]
\small
\centering
    \caption{Examples of tone-converted descriptions. The descriptions are written by an item supplier in the hotel domain, from which a specific sentence is extracted and converted in tone.}
    \label{tb:tone_example}
    \begin{tabular}{lp{14.5cm}}
     \toprule
    Tone & Descriptions\\ \midrule
    Original & Each room is a private villa, tastefully decorated to complement the surrounding nature, offering an ultra-luxurious stay that you won't find anywhere else.\\
    Neutral & The rooms, designed as private villas, have decor that harmonizes with the natural surroundings and offer a unique luxury experience.\\
    Attractive & Each room, a private villa, is elegantly adorned to mirror the natural beauty surrounding it, offering an unrivaled, lavish stay. Discover the allure of seclusion and class at our sanctuary. Highlight: Unparalleled Privacy and Luxurious Rooms.\\
    Humorous & Each room? More like a private villa, dressed to impress Mother Nature herself. It's the kind of luxury that makes Buckingham Palace look like a budget motel. \\
    Romantic & Each room is a private villa, decorated with a gentle touch to mirror the beauty of nature that surrounds us, offering a stay steeped in luxury and romance that is unparalleled. \\
    Formal & Each accommodation is an individual villa, elegantly appointed to harmonize with the surrounding natural environment, providing an unparalleled luxury stay unlike any other. \\
    Simple & Each room is a private villa, decorated to match the natural surroundings. It's a very luxurious stay that's unique to us. \\
    \bottomrule
    \end{tabular}
\end{table*}

\subsection{Measurement of Each Metric (Perceptive Effects of Tones)}\label{subsec:metric}
To investigate the impact of tones on user perception, we focused on seven explanation goals and the original metric, \textit{appeal}, as mentioned in the introduction.
Among the explanation goals, trust was defined by factors including \textit{competence}, \textit{benevolence}, and \textit{integrity}~\cite{wang2007recommendation}.
Therefore, we prepared a total of 10 different metrics. We used the questionnaire items as listed in Table~\ref{tb:questions}, referring to the previous study~\cite{Guesmi2022detail}.

As the method of ranking all tones for each metric burdened participants and did not capture the subtle differences between tones, we employed a pairwise comparison of tones. We asked the participants to compare tones for each questionnaire item and respond using a 7-point Likert scale. The method allowed us to collect scores reflecting how participants perceived each metric through pairwise comparison.

\begin{table*}[t]
\centering
\caption{Overview of survey questionnaire items.}
\label{tb:questions}
\small
\begin{tabular}{ll}
\toprule
Metric & Statement (Which description $\dots$)\\ \midrule
Transparency (Trans.) & helps you better understand what the recommendations are based on?\\
Scrutability (Scrut.) & better allows you to give feedback on how well your preferences have been understood?\\
Trust (Competence (Comp.)) & shows that the system has the expertise to understand your needs and preferences?\\
Trust (Benevolence (Benev.)) & shows that the system keeps your interests in mind?\\
Trust (Integrity (Integ.)) & shows that the system is honest? \\
Effectiveness (Effec.) & helps you determine how well the recommendations match your interests?\\ 
Persuasiveness (Pers.) & is more convincing?\\
Efficiency (Effi.) & helps you determine faster how well the recommendations match your interests?\\
Satisfaction (Satis.) & do you think is better?\\
Appeal (App.) & arouses your interest in the item more?\\
\bottomrule
\end{tabular}
\end{table*}

\subsection{Procedure}\label{subsec:procedure}

We conducted an online user study to answer the aforementioned research questions using the created dataset mentioned in Sec.~\ref{subsec:data}.
For this user study, we recruited participants through Amazon Mechanical Turk (MTurk) and asked them to perform pairwise comparisons of descriptions in a web browser.

In MTurk, each task is called a Human Intelligence Task (HIT).
We prepared three HITs corresponding to each domain, and each HIT was in the same format across the domains. We asked participants to assume they used a recommender system for the target domain. 
All HITs were conducted in English.
We estimated that each HIT would take approximately 30 min. to complete and offered a reward of 4.5 USD.
The details of each HIT are as follows.

\begin{enumerate}
    \setlength{\leftskip}{-0.5cm}
    \item The participants answered questions about their attributes (gender and age) and Big Five personality traits~\cite{john1999big}, which were estimated using a simple test proposed in the literature~\cite{gosling2003very}.
    \item The participants were presented with one recommended item and two descriptions. 
    The two descriptions were in different tones converted based on the same original description. They were labeled as descriptions A and B.
    \item The participants answered the 10 questions listed in Table~\ref{tb:questions} with a 7-point Likert scale (A is better - neither - B is better). We randomly displayed the 10 questions to avoid position bias.
    \item The participants repeated steps (2) and (3) mentioned above a total of 15 times, once for each of the 15 items. We assigned one of 15 items to each of the 15 ($={}_6 \mathrm{C}_2$) combinations of tones in a one-to-one manner, ensuring that participants evaluated all different tone pairs. 
    We randomized the order of tone pairs, tones in the pairs, and items to avoid bias.
\end{enumerate}

We designed the above procedure to focus on the impact of description tone on user perception.
Moreover, we used random items as recommended items to mitigate the influence of prediction accuracy of preferences.

\subsection{Participants and Demographic Information}\label{subsec:participant}

We recruited 300 participants per HIT (i.e., domain).
To maintain participant quality, we required the participants to possess an approval rate of at least 97\% on previous HITs and to have completed more than 50 HITs.
Additionally, we restricted the participants to residents in the US, UK, or Canada.
Moreover, to maintain the quality of answers, we excluded the responses by low-quality participants who either failed an attention test or had a response time under 15 min.
Therefore, the number of participants included 151 (movies), 170 (hotels), and 149 (products).
Table~\ref{tb:participants} shows the demographic information for each domain.
We collected participant responses from those with various attributes.

\begin{table}[t]
\centering
\small
\caption{Demographic information}
\label{tb:participants}
\begin{tabular}{@{}p{0.8cm}p{1.8cm}p{1.2cm}p{3.5cm}@{}}
\toprule
Domain & \# of participants & Gender & Mean Age\\ 
\midrule
Movies  & 151 & male: 105, female: 46 & 36.2 (SD: 10.1, range: 23 --76)  \\
Hotels  & 170 & male: 104, female: 66 & 35.3 (SD: 9.4, range: 23 --75) \\
Products & 149 & male: 81, female: 68 & 35.5 (SD: 9.9, range: 21 --78) \\
\bottomrule
\end{tabular}
\end{table}

Big Five personality comprised five personality traits: extroversion, agreeableness, conscientiousness, neuroticism, and openness.
We calculated the score of each trait based on the collected responses according to the literature~\cite{gosling2003very} and then normalized the score from 0 to 1.
Table~\ref{tb:personality} shows the characteristics of each trait and the distribution of participants.
We collected participants with similar distributions of personality traits in all domains.

\begin{table*}[t]
\centering
\small
\caption{Distribution of Big Five personality traits among participants. Each value represents the mean value for each trait, and the values in $(\cdot)$ represent the standard deviation.}
\label{tb:personality}
\begin{tabular}{@{}llccc@{}}
\toprule
Personality Trait & Questionnaire Items (I see myself as ...) & Movies & Hotels & Products \\ 
\midrule
Extroversion (Ex.)      &  Extroverted, enthusiastic. / Reserved, quiet. & 0.51 (0.17) & 0.52 (0.19) & 0.53 (0.17) \\
Agreeableness (Ag.) & Sympathetic, warm. / Critical, quarrelsome.  & 0.63 (0.19) & 0.65 (0.17) & 0.64 (0.17) \\
Conscientiousness (Co.) & Dependable, self-disciplined. / Disorganized, careless. & 0.64 (0.21) & 0.65 (0.21) & 0.67 (0.14) \\
Neuroticism (Ne.) &  Anxious, easily upset./ Calm, emotionally stable. & 0.45 (0.19) & 0.43 (0.18) & 0.42 (0.19) \\
Openness (Op.) & Open to new experiences, complex. / Conventional, uncreative. & 0.63 (0.19) & 0.62 (0.18) & 0.61 (0.18) \\
\bottomrule
\end{tabular}
\end{table*}

\subsection{Ethical Concern}\label{subsec:ethic}
Before starting the user study, the participants were informed about this survey, risks, and rights.
They were asked for consent to use the obtained user data and responses in a non-identifiable form for research purposes.
We conducted the user study only for consenting participants.
The participants were allowed to stop the user study at any time.
We collected worker IDs on MTurk for identifying each user, but the information was only used to pay participation rewards.
Moreover, we used the collected data only for research purposes and did not share it with any other party to preserve the participants' privacy.
Our institutional ethical committee approved the study protocol (approval ID: 22B-08).

\section{Results and Discussions}\label{sec:result}
In this section, we first explain the methods used for analyzing the collected data.
Next, we present the results of average tonal effects that do not depend on user attributes, and then discuss RQs 1 and 2.
Further, we present the results of tonal effects that are dependent on user attributes, and discuss RQ 3.

\subsection{Analysis Method}\label{subsec:analysis}
We conducted an ordinal logistic regression analysis to investigate how the tone of descriptions influences each metric.
The method is particularly suited for modeling Likert scale data from pairwise comparisons.
The collected data comprised scores derived from the pairwise comparison of two descriptions with different tones (descriptions A and B) for each metric.
The descriptions, based on the same original description, were expressed in different tones, namely tones A and B. 
Thus, the collected scores can be seen as direct comparative evaluation values of tone A against tone B.

We denoted a set of user attributes $\mathcal{A}$ and a set of tones $\mathcal{T}$ as follows.
\begin{align}
    \mathcal{A} &= \{\text{Age, Gender, Ex., Ag., Co., Ne., Op.}\},\nonumber \\ 
    \mathcal{T} &= \{\mathrm{Neu.}, \mathrm{Att.}, \mathrm{Hum.}, \mathrm{Rom.}, \mathrm{For.}, \mathrm{Sim.}\}. \nonumber
\end{align}
We defined the utility of tone $t \in \mathcal{T}$ for each metric as $u_{t} := \beta_{t} + \sum_{e \in \mathcal{A}} \beta_{t, e} X_{e}$, where $\beta_t$ and $\beta_{t, e}, e \in \mathcal{A}$ represent the coefficients of the logistic model, and $X_{e}, e \in \mathcal{A}$ represents independent variables. 
When we denoted $t_A$ and $t_B$ as tones A and B, respectively, we modeled the scores of tone A over those of tone B as the difference between utilities: $u_{t_A} - u_{t_B}$.
We fitted the ordinal logistic regression model with the scores from pairwise comparisons as a dependent variable $S_i \in \{-3, -2, -1, 0, +1, +2, +3\} =: \mathcal{J}$, treating them as ordinal categorical scores, where $i$ represents the index of collected data.
We used six tones and user attributes as independent variables, constructing the model with terms that included each tone and its interactions with user attributes (age, gender, and personality traits).
In order to model the scores $S_i$ of comparisons between tones A and B, we rewrote $u_t$ using dummy variable $X_t$ and constructed the ordinal logistic regression model as follows:
\begin{align}
    \text{logit} (P(S_i \leq j)) &= \gamma_j - (u_{t_A} - u_{t_B}) \nonumber \\
    &= \gamma_j - \sum_{t \in T} \sum_{e \in \mathcal{A}} (\beta_t X_t + \beta_{t, e} X_t X_{e}),
\end{align}
where $\text{logit}(p)=\log (p/(1-p))$ is the logit function, $P(S_i \leq j)$ represents the cumulative probability of $S_i$ less than or equal to a specific category $j \in \mathcal{J}$, $\gamma_j$ is the threshold parameter for category $j \in \mathcal{J}$.
For the dummy variable $X_t$, we assigned 1 when it corresponded to tone A, -1 on corresponding to tone B, and 0 otherwise.
Gender was treated as a dummy variable, while age and personality traits were normalized to have a mean of 0 and a variance of 1.

For our analysis, we constructed the ordinal logistic regression model with reference to a neutral tone and fitted the collected data to it.
For each independent variable, the coefficient was tested under the null hypothesis that the coefficient was 0, using a significance level of $\alpha = 0.05$.
When the coefficient was positive and the null hypothesis was rejected, it indicated that the independent variable had a positive impact compared to the neutral tone.
We conducted the analysis across each domain and applied false discovery rate correction following Benjamini--Hochberg procedure to address the issue of multiple comparisons.
We implemented our analyses by Python and used \textit{statsmodels} library~\cite{seabold2010statsmodels}.

\subsection{General impact of tones regardless of user attributes}\label{subsec:r1}
First, we focused on the impact of each tone for an average participant by removing the influence of user attributes (age, gender, and personality traits).
Evaluation was based on an analysis of intercept terms $\beta_t$ for each $t \in \mathcal{T}$ of the model.
Table~\ref{tb:r1} presents the intercepts with values in parentheses representing p-values.
The values where the null hypothesis was rejected are highlighted in bold.

\begin{table*}[t]
\caption{Results of testing the average effects of tones, removing the influence of user attributes. The values represent coefficients, with values in parentheses representing p-values. Bold values show the rejection of null hypothesis.} 
\label{tb:r1}
\footnotesize
\centering
\textbf{Movies} \\
\begin{tabular}{@{}p{4.5mm}p{12.0mm}p{12.0mm}p{12.0mm}p{12.0mm}p{12.0mm}p{12.0mm}p{12.0mm}p{12.0mm}p{12.0mm}p{12.0mm}@{}}
\toprule
 Tone & Trans. & Scrut. & Comp. & Benev. & Integ. & Effec. & Pers. & Effi. & Satis. & App. \\ \midrule
Att. & $0.27 (0.36)$ & $0.44 (0.05)$  & $0.26 (0.37)$ & $0.21 (0.49)$ & $0.30 (0.31)$ & $\textbf{0.48 (<.05)}$ & $0.24 (0.42)$ & $0.20 (0.53)$ & $0.38 (0.17)$ & $0.19 (0.53)$ \\ 
Hum. & $-0.13 (0.70)$ & $-0.09 (0.80)$ & $-0.11 (0.75)$ & $-0.15 (0.65)$ & $-0.10 (0.78)$ & $-0.13 (0.69)$ & $-0.27 (0.36)$ & $-0.09 (0.80)$ & $-0.25 (0.40)$ & $-0.09 (0.80)$ \\ 
Rom. & $0.04 (0.89)$ & $0.30 (0.30)$  & $0.18 (0.58)$ & $0.12 (0.71)$ & $0.09 (0.80)$ & $0.23 (0.43)$ & $0.07 (0.84)$ & $-0.07 (0.85)$ & $0.22 (0.46)$ & $-0.04 (0.91)$ \\ 
For. & $0.05 (0.87)$ & $0.21 (0.48)$  & $0.20 (0.51)$ & $0.04 (0.90)$ & $0.04 (0.90)$ & $0.22 (0.46)$ & $0.03 (0.93)$ & $0.02 (0.96)$ & $0.16 (0.62)$ & $0.01 (0.96)$ \\ 
Sim. & $-0.27 (0.36)$ & $-0.07 (0.84)$ & $-0.03 (0.95)$ & $-0.31 (0.29)$ & $-0.15 (0.66)$ & $-0.05 (0.87)$ & $-0.22 (0.44)$ & $-0.21 (0.49)$ & $-0.16 (0.62)$ & $-0.23 (0.44)$ \\ 

\bottomrule
\end{tabular} \\
\textbf{Hotels} \\
\begin{tabular}{@{}p{4.5mm}p{12.0mm}p{12.0mm}p{12.0mm}p{12.0mm}p{12.0mm}p{12.0mm}p{12.0mm}p{12.0mm}p{12.0mm}p{12.0mm}@{}}
\toprule
Att. & $\textbf{0.38 (<.05)}$ & $0.32 (0.11)$ & $\textbf{0.40 (<.05)}$ & $\textbf{0.53 (<.05)}$ & $0.14 (0.59)$ & $\textbf{0.42 (<.05)}$ & $0.33 (0.09)$ & $\textbf{0.44 (<.05)}$ & $0.23 (0.30)$ & $\textbf{0.38 (<.05)}$ \\ 
Hum. & $0.15 (0.56)$ & $0.14 (0.60)$  & $0.17 (0.46)$ & $0.17 (0.46)$ & $-0.09 (0.74)$ & $0.33 (0.09)$ & $0.18 (0.44)$ & $0.18 (0.45)$ & $-0.12 (0.63)$ & $0.24 (0.27)$ \\ 
Rom. & $0.31 (0.12)$ & $\textbf{0.47 (<.05)}$ & $0.32 (0.09)$ & $\textbf{0.55 (<.05)}$ & $0.13 (0.64)$ & $\textbf{0.53 (<.05)}$ & $\textbf{0.37 (<.05)}$ & $\textbf{0.38 (<.05)}$ & $0.24 (0.28)$ & $\textbf{0.40 (<.05)}$ \\ 
For. & $\textbf{0.52 (<.05)}$ & $\textbf{0.48 (<.05)}$ & $0.21 (0.33)$ & $\textbf{0.42 (<.05)}$ & $0.10 (0.71)$ & $\textbf{0.46 (<.05)}$ & $\textbf{0.42 (<.05)}$ & $\textbf{0.51 (<.05)}$ & $\textbf{0.36 (<.05)}$ & $\textbf{0.44 (<.05)}$ \\ 
Sim. & $-0.02 (0.97)$ & $0.09 (0.77)$ & $0.06 (0.86)$ & $0.01 (0.99)$ & $-0.08 (0.79)$ & $0.10 (0.71)$ & $-0.02 (0.97)$ & $0.07 (0.81)$ & $-0.19 (0.41)$ & $-0.16 (0.51)$ \\ 

\bottomrule
\end{tabular} \\
\textbf{Products} \\
\begin{tabular}{@{}p{4.5mm}p{12.0mm}p{12.0mm}p{12.0mm}p{12.0mm}p{12.0mm}p{12.0mm}p{12.0mm}p{12.0mm}p{12.0mm}p{12.0mm}@{}}
\toprule
Att. & $0.35 (0.11)$ & $0.24 (0.29)$  & $0.29 (0.19)$ & $0.21 (0.38)$ & $0.17 (0.50)$ & $0.30 (0.16)$ & $0.41 (0.05)$ & $\textbf{0.42 (<.05)}$ & $\textbf{0.42 (<.05)}$ & $0.31 (0.16)$ \\ 
Hum. & $0.03 (0.94)$ & $0.13 (0.64)$  & $0.04 (0.93)$ & $-0.01 (0.97)$ & $-0.10 (0.73)$ & $0.03 (0.94)$ & $0.22 (0.34)$ & $-0.07 (0.87)$ & $-0.12 (0.69)$ & $-0.04 (0.92)$ \\ 
Rom. & $0.37 (0.08)$ & $0.38 (0.07)$  & $0.31 (0.16)$ & $0.31 (0.16)$ & $0.25 (0.27)$ & $0.27 (0.23)$ & $\textbf{0.49 (<.05)}$ & $0.37 (0.08)$ & $0.17 (0.53)$ & $0.32 (0.14)$ \\ 
For. & $0.36 (0.09)$ & $0.41 (0.05)$  & $0.30 (0.18)$ & $0.25 (0.28)$ & $0.24 (0.29)$ & $0.22 (0.34)$ & $\textbf{0.44 (<.05)}$ & $0.26 (0.27)$ & $0.38 (0.08)$ & $0.26 (0.26)$ \\ 
Sim. & $-0.07 (0.85)$ & $-0.22 (0.32)$ & $-0.05 (0.91)$ & $-0.23 (0.32)$ & $-0.22 (0.32)$ & $-0.30 (0.16)$ & $-0.04 (0.92)$ & $-0.09 (0.78)$ & $-0.10 (0.72)$ & $-0.36 (0.08)$ \\ 

\bottomrule
\end{tabular}
\end{table*}

\subsubsection{RQ1. How does the tone of explanations influence perceived effects regardless of domains and user attributes?}

Table~\ref{tb:r1} indicates that tones influenced every metric excluding integrity in at least one domain.
Notably, although the metrics, such as scrutability and competence, relate to the perceived mechanism of the recommender system and were expected to be judged solely by the information contained in the description, they are influenced by tones as well.
The results suggest that tones may be crucial in how users process and interpret information in the descriptions.

Furthermore, a common trend observed across domains involves no significant difference in simple tone compared to neutral tone. 
Conversely, both attractive and romantic tones positively influence many metrics compared to the neutral tone.
Thus, descriptions with rich and varied expressions generally tend to influence users more positively across various metrics than plain and uniform descriptions.

\subsubsection{RQ2. How do the perceived effects of tones vary across domains regardless of user attributes?}

We hypothesized that tone would influence domains with entertainment elements more than practical domains owing to a lower risk of item consumption.
However, the results differed; tone influenced almost all metrics in the hotel domain, while only effectiveness was influenced in the movie domain.
Notably, significant influences were observed for appeal in the hotel domain.
The results are likely due to users' different expectations for the descriptions depending on the domains.
In the movie domain, users may focus on whether the content aligns with their preferences.
In contrast, in the hotel domain, emotional elements such as evoking a relaxed mood or enjoyable experience may be valued, along with accuracy and concreteness of information.

Moreover, we expected the influences of tone to be limited in practical domains, including products.
However, significant influences of tones were observed in persuasiveness, effectiveness, and satisfaction.
This might be because that tone supports users' imaginations of their experiences with products, leading to the impacts on these metrics. 
These results imply that tone can make product items look fascinating and stimulate purchasing desires.

Focusing on the type of tone, we found a significant effect of formal tones in hotel and product domains.
In domains closer to practical applications like hotels and products, users would prefer professional, specific, and accurate information, where the formal tone aligns well with the expectations. 
In the movie domain, users may seek narrative richness and emotional engagement, which can be hard to convey through the formal tone.

In summary, the impacts of tones vary across domains.
Therefore, appropriate adjustment of description tone according to domains can maximize the perceived effects of recommender systems.

\subsection{Impact of tones depending on user attributes}\label{subsec:r2}

The results presented in Table~\ref{tb:r1} reflect the general influence of each tone, independent of user attributes.
Thus, for metrics without significant differences, the effects of tone can vary among different users or be non-existent.
Consequently, we further investigated the impacts of interactions between each tone and user attributes.

Table~\ref{tb:r2} presents the coefficients of interaction terms between each tone and each user attribute for each metric, with values in parentheses representing p-values.
The values where the null hypothesis was rejected are highlighted in bold, indicating significant influences compared to a neutral tone.
Due to space limitations, we have only included the results of interaction terms with significant differences.

\begin{table*}[t]
\footnotesize
\caption{Results of testing the influence of interaction between each tone and each user attribute.
The values represent coefficients, with values in parentheses representing p-values. Bold values indicate rejection of null hypothesis.} 
\label{tb:r2}
\centering
\textbf{Movies} \\
\begin{tabular}{@{}p{4.0mm}p{4.5mm}p{12.0mm}p{12.0mm}p{12.0mm}p{12.0mm}p{12.0mm}p{12.0mm}p{12.0mm}p{12.0mm}p{12.0mm}p{12.0mm}@{}}
\toprule
Attri. & Tone & Trans. & Scrut. & Comp. & Benev. & Integ. & Effec. & Pers. & Effi. & Satis. & App. \\ \midrule
Ex. & Att. & $\textbf{0.28 (<.05)}$ & $\textbf{0.28 (<.05)}$ & $\textbf{0.44 (<.05)}$ & $\textbf{0.39 (<.05)}$ & $0.20 (0.22)$ & $\textbf{0.31 (<.05)}$ & $0.22 (0.18)$ & $0.22 (0.18)$ & $0.23 (0.16)$ & $0.22 (0.18)$ \\ 
Ex. & Hum. & $\textbf{0.35 (<.05)}$ & $\textbf{0.30 (<.05)}$ & $\textbf{0.34 (<.05)}$ & $\textbf{0.32 (<.05)}$ & $\textbf{0.28 (<.05)}$ & $\textbf{0.32 (<.05)}$ & $0.26 (0.07)$ & $0.17 (0.35)$ & $0.28 (0.05)$ & $0.21 (0.22)$ \\ 
Ex. & For. & $\textbf{0.30 (<.05)}$ & $\textbf{0.28 (<.05)}$ & $\textbf{0.41 (<.05)}$ & $\textbf{0.40 (<.05)}$ & $0.25 (0.09)$ & $\textbf{0.38 (<.05)}$ & $0.24 (0.13)$ & $0.20 (0.23)$ & $0.18 (0.29)$ & $0.12 (0.49)$ \\ 
Co. & Rom. & $0.25 (0.24)$ & $0.10 (0.71)$ & $\textbf{0.35 (<.05)}$ & $0.12 (0.63)$ & $0.05 (0.85)$ & $0.13 (0.59)$ & $0.18 (0.43)$ & $0.10 (0.69)$ & $0.02 (0.95)$ & $0.32 (0.09)$ \\ 
Ne. & For. & $-0.21 (0.35)$ & $-0.14 (0.56)$ & $-0.09 (0.71)$ & $-0.18 (0.42)$ & $-0.20 (0.36)$ & $-0.04 (0.87)$ & $-0.11 (0.66)$ & $-0.23 (0.27)$ & $\textbf{-0.39 (<.05)}$ & $-0.09 (0.71)$ \\ 
\bottomrule
\end{tabular} \\

\textbf{Hotels} \\ 
\begin{tabular}{@{}p{4.0mm}p{4.5mm}p{12.0mm}p{12.0mm}p{12.0mm}p{12.0mm}p{12.0mm}p{12.0mm}p{12.0mm}p{12.0mm}p{12.0mm}p{12.0mm}@{}}
\toprule
Age & Att. & $-0.04 (0.82)$ & $-0.12 (0.39)$ & $-0.11 (0.44)$ & $-0.01 (0.97)$ & $\textbf{-0.30 (<.05)}$ & $-0.07 (0.70)$ & $-0.17 (0.21)$ & $-0.14 (0.32)$ & $-0.11 (0.46)$ & $0.01 (0.97)$ \\ 
Ex. & Att. & $\textbf{0.39 (<.05)}$ & $\textbf{0.24 (<.05)}$ & $0.19 (0.12)$ & $0.13 (0.33)$ & $\textbf{0.23 (<.05)}$ & $0.16 (0.23)$ & $\textbf{0.28 (<.05)}$ & $\textbf{0.22 (<.05)}$ & $\textbf{0.37 (<.05)}$ & $\textbf{0.34 (<.05)}$ \\ 
Ex. & Hum. & $\textbf{0.27 (<.05)}$ & $\textbf{0.22 (<.05)}$ & $\textbf{0.30 (<.05)}$ & $\textbf{0.29 (<.05)}$ & $\textbf{0.29 (<.05)}$ & $\textbf{0.27 (<.05)}$ & $\textbf{0.34 (<.05)}$ & $\textbf{0.23 (<.05)}$ & $\textbf{0.31 (<.05)}$ & $0.22 (0.07)$ \\ 
Ex. & Rom. & $\textbf{0.30 (<.05)}$ & $0.16 (0.21)$ & $0.22 (0.05)$ & $\textbf{0.24 (<.05)}$ & $\textbf{0.29 (<.05)}$ & 0.22 $(0.05)$ & $\textbf{0.26 (<.05)}$ & $0.20 (0.10)$ & $\textbf{0.34 (<.05)}$ & $\textbf{0.33 (<.05)}$ \\ 

\bottomrule
\end{tabular}
\\

\textbf{Products} \\
\begin{tabular}{@{}p{4.0mm}p{4.5mm}p{12.0mm}p{12.0mm}p{12.0mm}p{12.0mm}p{12.0mm}p{12.0mm}p{12.0mm}p{12.0mm}p{12.0mm}p{12.0mm}@{}}
\toprule
Age & Hum. & $-0.20 (0.20)$ & $-0.12 (0.47)$ & $-0.20 (0.18)$ & $-0.22 (0.16)$ & $-0.16 (0.32)$ & $-0.16 (0.32)$ & $-0.24 (0.12)$ & $-0.29 (0.07)$ & $\textbf{-0.35 (<.05)}$ & $-0.30 (0.05)$ \\ 
Age & Rom. & $-0.23 (0.11)$ & $-0.16 (0.29)$  & $-0.25 (0.08)$ & $-0.25 (0.09)$ & $-0.16 (0.31)$ & $-0.28 (0.05)$ & $\textbf{-0.29 (<.05)}$ & $\textbf{-0.31 (<.05)}$ & $\textbf{-0.29 (<.05)}$ & $\textbf{-0.30 (<.05)}$ \\

\bottomrule
\end{tabular}
\end{table*}

\subsubsection{RQ3. How do the perceived effects of tones depend on user attributes?}

From Table~\ref{tb:r2}, participants with high extroversion are found to be more influenced by tones in movie and hotel domains.
Given that extroverted people tend to be cheerful and enjoy communication with others, it is conceivable that they responded more positively to emotionally rich tones, such as attractive, humorous, and romantic, that are more human-like and approachable across multiple metrics. 
Notably, in the movie domain, where the average impact of tone is limited (Table~\ref{tb:r1}), participants with high extroversion are influenced by tones in metrics related to the recommender system mechanism perception, including transparency, scrutability, and trust (competence, benevolence, and integrity).
The results may reflect the tendency of highly extroverted participants to have clear movie preferences and try to understand the rationale for their recommendations.
Conversely, extroversion shows no significant effect in the product domain. This could be because, in contrast to domains focused on entertainment and experience, such as movies and hotels, the product domain emphasized practical or functional aspects. As a result, there was less of an emotional component, even for participants with high extroversion.

For other personality traits, participants with higher conscientiousness show positive impact on competence by the romantic tone in the movie domain.
Additionally, those with higher neuroticism tend to perceive satisfaction negatively when exposed to a formal tone. 
Thus, the impacts of tones depend on personalities.
Moreover, elderly participants tend to respond negatively to attractive, humorous, and romantic tones.
This trend implies there are tone sensitivity differences associated with age, indicating elderly people may prefer more conservative, information-based tones.

In summary, the perceived effects of tones vary with user attributes.
These results suggested that adjusting the tone of descriptions to consider user personality traits and age could enhance the perceived effects of descriptions in recommender systems.

\subsection{Limitation}\label{subsec:limit}
We explored the impact of description tones by analyzing data collected through user experiments, focusing on three aspects: perceived effects, domains, and user attributes.
Our findings reveal that tone significantly influences various metrics and tonal effects vary with certain user attributes. However, our study has the following limitations:\\
\textbf{Fictitious data source}: 
We utilized fictitious data sources generated using LLM to prepare a dataset suited for answering our research questions, as existing data sources were inadequate. While we have detailed the data source generation procedure, the steps involve randomness associated with LLMs and may not be completely replicable. Furthermore, the generated data sources may not fully reflect the descriptions found in actual services, potentially limiting the generalizability of our results.\\
\textbf{Tone selection}: This study used only six types of tones and treated each as a distinct category. However, tones can be more diverse, with varying intensities and types, forming a continuous spectrum. Therefore, the range of tones covered in this study may not comprehensively capture the full impact of tone.\\
\textbf{Change in perceived content by tone}: 
Our user study asked participants to compare tones of the same description for each metric. This approach assumes that only tone differs while the description contents remain the same. However, tone actually influences the perception of contents (Appendix~\ref{app:gpt_survey}). Complete separation of tone from content is challenging, and the results may have been influenced by the perceived differences in content for each metric.
\\
\textbf{Limitation on user attributes}: 
This study considers age, gender, and the Big Five personality traits as user attributes. However, cultural background and expertise are known to influence the perception of descriptions~\cite{knijnenburg2012explaining}. Not accounting for these factors may limit the generalization of our research findings.\\
Such limitations highlight directions for future research, including the use of more realistic data sources, expansion of tone spectra, and consideration of a broader range of user attributes.

\section{Conclusions}\label{sec:conc}
This study investigated the impact of the tone of explanations (i.e., item descriptions) in recommender systems from three perspectives:
(1) perceived effects of tone, (2) relation between the effects of tones and domain, and (3) relation between the effects of tones and user attributes.
For this study, we generated fictitious data sources using LLM, including descriptions across movie, hotel, and home product domains.
Further, we converted the descriptions into those with six tones.
In our online user study, participants assessed several metrics through pairwise comparisons of the six tones.
We analyzed the collected data using an ordinal logistic regression model to investigate the influence of tones on various metrics and their variations due to user attributes across three domains. 
(1) The results showed significant influences across almost all perceived effects, such as transparency, effectiveness, and persuasiveness.
(2) Particularly, the effects of tones were observed in the hotel domain more than in the movie and product domains.
(3) We found that the effects of tones were not uniform across all users; they varied with user attributes, including age and personality traits.

Our findings suggest that an appropriate adjustment of description tones according to domains and user attributes can maximize the perceived effects of recommender systems.
However, a potential risk of these influences leading to users' overestimation exists; therefore, careful consideration is required when incorporating such tones in descriptions.
Based on our findings, we plan to develop methods that predict the most effective tone for individuals.

\appendix
\section*{Appendix}
\section{Perceptive changes in content due to tone conversion}
\label{app:gpt_survey}

\begin{table*}[th]
    \centering
    % \small
    \caption{Mean values for each tone, considering information consistency and level of detail, compared to the original description across domains. Values in parentheses represent standard deviations. (1) Values correspond to the 5-point Likert scale. (2) Value 0 indicates similar detail as the original description; positive values indicate higher detail; and negative values indicate lower detail.}
    \label{tb:gpt_check}
    \begin{tabular}{lcccccc}   
     & Neutral & Attractive & Humorous & Romantic & Formal & Simple\\ \hline
    (1) information consistency & 2.77 (1.02) & 2.76 (1.00) & 2.83 (1.00) & 2.83 (0.98) & 2.78 (0.97) & 2.66 (0.97) \\
    (2) detail & 0.20 (1.11) & 0.53 (1.14) & 0.47 (1.17) & 0.56 (1.23) & 0.34 (1.12) & 0.08 (1.15) \\
    \bottomrule
    \end{tabular}
\end{table*}

In our user study described in Sec.~\ref{sec:user_study}, we asked participants to compare tones for the same descriptions.
The setting assumed that the descriptions had the same amount of information (i.e., the same content) even after converting the same descriptions into different tones using LLM.
However, the change in tone might cause participants to perceive the content differently.
Therefore, we verified whether the conversion of tones led to any perceptive changes in the description contents for the following aspects: information consistency and level of detail.

\paragraph{Setting}
Our survey was conducted on the web using Amazon Mechanical Turk. 
We recruited 100 participants per domain, with the same requirements as the user study in Sec.~\ref{sec:user_study}.
Each HIT was estimated to take 20 min., with a compensation of 3.0 USD.
After removing low-quality participants who failed an attention test, we obtained 50 (movies), 46 (hotels), and 43 (products) responses.
We randomly presented participants with 12 items and a pair of original descriptions (description A) and tone-converted descriptions (description B).
Each tone was randomly assigned to description B, ensuring each participant responded to each tone twice.
The participants answered the following questions:
\begin{enumerate}
    \item \textit{Do Description A and Description B contain the same information despite differences in style and expression?} (5-point Likert scale. 1: exactly the same, 2: mostly the same information with very minor differences, 3: mostly the some information, but with some differences, 4: significant differences, 5: completely different)
    \item \textit{How does the level of detail in Description B compare to Description A?} (5-point Likert scale.)
\end{enumerate}

\paragraph{Results}

Table~\ref{tb:gpt_check} presents the mean and standard deviation results for responses to each tone across all domains.
We conducted an ANOVA analysis at a 5\% significance level.
The ANOVA analysis showed that while the null hypothesis was not rejected for information consistency ($F=1.35$, $p=0.25$), it was rejected for the level of detail ($F=7.82$, $p<.05$). Regarding information consistency, many participants answered the tone-converted descriptions as (2) \textit{mostly the same information with very minor differences} or (3) \textit{mostly the some information, but with some differences}.
Focusing on the level of detail, many participants perceived emotional tones, such as attractive, humorous, and romantic, to be more detailed compared to the original descriptions.
The results suggested that perceived content was also influenced by tone conversion.
Therefore, the main user survey's results may have been influenced by the perceived differences in content for each metric.

\bibliographystyle{ACM-Reference-Format}
\bibliography{reference}

\end{document}